\documentclass[aps,twocolumn,showpacs]{revtex4}
\usepackage{epsfig}

\newcommand{\zaa}{Astron.~Astrophys.}

\newcommand{\zapj}{Astrophys.~J.}

\newcommand{\zapjs}{Astrophys.~J.~Suppl. Ser.}

\newcommand{\znp}{Nucl.~Phys.}

\newcommand{\zpr}{Phys.~Rev.}

\newcommand{\gap}{\mathrel{ \rlap{\raise.5ex\hbox{$>$}}
                    {\lower.5ex\hbox{$\sim$}}  } }
\newcommand{\lap}{\mathrel{ \rlap{\raise.5ex\hbox{$<$}}
                {\lower.5ex\hbox{$\sim$}}  } }
\newcommand{\ob}{$\Omega_b$}
\newcommand{\obh}{$\Omega_bh^2$}

\newcommand{\six}{$^{6}$Li}
\newcommand{\sep}{$^{7}$Li}

\newcommand{\zdag}{$^2$H$(\alpha,\gamma)^6$Li}

\begin{document}

\title{High-energy break-up of $^6$Li as a tool to study the Big-Bang nucleosynthesis reaction
$^2$H($\alpha$,$\gamma$)$^6$Li}

\author{
F.~{Hammache}$^{1,2}$, M.~Heil$^{2}$, S.~{Typel}$^{2,3}$,
D.~{Galaviz}$^{4}$\footnote{present address: Centro de F\'isica
Nuclear, Universidade de Lisboa, P-1649-003 Lisboa, Portugal},
K.~{S\"ummerer}$^2$\footnote{Electronic address:
k.suemmerer@gsi.de}, A.~{Coc}$^5$, F.~{Uhlig}$^2$,
F.~{Attallah}$^2$, M.~{Caamano}$^6$, D.~{Cortina}$^6$,
H.~{Geissel}$^2$, M.~{Hellstr\"om}$^2$, N.~{Iwasa}$^7$,
J.~{Kiener}$^5$, P.~{Koczon}$^2$, B.~{Kohlmeyer}$^8$,
P.~{Mohr}$^4$\footnote{present address: Diakonie-Klinikum
Schw\"abisch-Hall, D-74523 Schw\"abisch-Hall, Germany},
E.~{Schwab}$^2$, K.~{Schwarz}$^2$, F.~{Sch\"umann}$^9$,
P.~{Senger}$^2$, O.~{Sorlin}$^1$\footnote{present address: GANIL,
BP 55027, F-14076 Caen Cedex 5, France}, V.~{Tatischeff}$^5$,
J.P.~{Thibaud}$^5$, E.~{Vangioni}$^{10}$, A.~{Wagner}$^{11}$, and
W.~{Walus}$^{12}$ }
\affiliation{$^1$Institut de Physique Nucl\'eaire, UMR-8608,
CNRS/IN2P3 and Universit\'e Paris Sud XI, F-91406 Orsay, France}
\affiliation{$^2$GSI Helmholtzzentrum f\"ur Schwerionenforschung
GmbH, D-64220 Darmstadt, Germany}
\affiliation{$^{3}$Excellence Cluster ``Origin and Structure of
the Universe'', Technische Universit\"{a}t M\"{u}nchen, D-85748
Garching, Germany}
\affiliation{$^4$Technische Universit\"at Darmstadt, D-64220
Darmstadt, Germany}
\affiliation{$^5$Centre de Spectrom\'etrie Nucl\'eaire et de
Spectrom\'etrie de Masse (CSNSM), UMR-8609, CNRS/IN2P3 and
Universit\'e Paris Sud XI, F-91405 Orsay Campus, France}
\affiliation{$^6$Universidade Santiago de Compostela, E-15786
Santiago, Spain}
\affiliation{$^7$Tohoku University, Aoba, Sendai, Miyagi 980-8578,
Japan}
\affiliation{$^8$Fachbereich Physik, Philipps Universit\"at,
D-3550 Marburg, Germany}
\affiliation{$^9$Ruhr-Universit\"at Bochum, D-44780 Bochum,
Germany}
\affiliation{$^{10}$Institut d'Astrophysique de Paris, UMR-7095,
CNRS and Universit\'e Pierre et
 Marie Curie, F-75014 Paris, France}
\affiliation{$^{11}$Forschungszentrum Rossendorf, D-01314 Dresden,
Germany}
\affiliation{$^{12}$Institute of Physics, Jagiellonian University,
PL-30-059 Krakow, Poland}


\date{November 29, 2010}


\begin{abstract}
The recently claimed observations of non-negligible amounts of
$^6$Li in old halo stars have renewed interest in the Big-Bang
Nucleosynthesis (BBN) of $^6$Li. One important ingredient in the
predicted BBN abundance of \six\ is the low-energy
$^2$H($\alpha$,$\gamma$)$^6$Li cross section. Up to now, the only
available experimental result for this cross section showed an
almost constant astrophysical $S$-factor below 400 keV, contrary
to theoretical expectations. We report on a new measurement of the
$^2$H($\alpha$,$\gamma$)$^6$Li reaction using the break-up of
$^6$Li at 150 $A$ MeV. Even though we cannot separate
experimentally the Coulomb contribution from the nuclear one, we
find clear evidence for Coulomb-nuclear interference by analyzing
the scattering-angular distributions. This is in-line with our
theoretical description which indicates a drop of the
$S_{24}$-factor at low energies as predicted also by most other
models. Consequently, we find even lower upper limits for the
calculated primordial $^6$Li abundance than before.
\end{abstract}

\maketitle

\section{Introduction}

The Big--Bang model of the Universe is mainly supported by three
observational evidences: the expansion of the Universe, the Cosmic
Microwave Background (CMB), and the primordial or Big-Bang
Nucleosynthesis (BBN) of light nuclei like $^2$H, $^{3,4}$He, and
$^7$Li with one free parameter, the baryonic density of the
Universe, $\Omega_b$. Recently, a precise value for this free
parameter ($\Omega_b$h$^2$=0.02260$\pm$0.00053) has been deduced
from the a\-na\-ly\-sis of the anisotropies in the CMB as observed
by the WMAP satellite \cite{WMAP}. Using the best available
nuclear reaction rates, this now allows to make precise
predictions for the primordial abundances of these light nuclei. A
com\-pa\-ri\-son between calculations and observations reveals
good agreement for helium, excellent agreement for deuterium, and
a discrepancy (by a factor of $\approx4$) for
\sep~\cite{Coc04,Cyb08,Coc10}. Possible reasons for this
discrepancy for \sep\ have recently been discussed e.g. by Spite
and Spite~\cite{Spi10}.

In 2006, high-resolution observations of Li absorption lines in
some very old halo stars have led the authors to claim evidence
for large primordial abundances also of the weakly-bound isotope
$^6$Li \cite{Asp06}. The \six/\sep\ ratios of $\sim 5\times
10^{-2}$ were found to be about three orders of magnitude larger
than the BBN-calculated value of \six/\sep~$\sim 10^{-5}$. This
observation has triggered many studies to resolve the discrepancy
either by considering an early $^6$Li formation in primitive dwarf
galaxies at high redshift in a hierarchical-structure formation
context~\cite{Rol06}, in situ by solar-like flares~\cite{TT07}, or
in terms of physics beyond the standard model of particle physics
(see, e.g., Refs.~\cite{Kus07,Eli05,Jed06}). More recently,
however, Cayrel {\it et al.}~\cite{Cay07} and Steffen {\it et
al.}~\cite{Ste10} have pointed out that line asymmetries similar
to those created by a $^6$Li blend could also be produced by
convective Doppler shifts in stellar atmospheres. Similarly, a
recent study of Garcia Perez {\it et al.}~\cite{Gar09} could not
claim any significant detection of $^6$Li in metal-poor stars. So,
presently the debate is open. More stellar observations are
required to solve this question (see Asplund and Lind
~\cite{Asp10}).

Predictions for the production of \six\ in BBN require precise
measurements of the $^2$H($\alpha$,$\gamma$)$^6$Li reaction rate,
the key production mechanism. In BBN, this reaction occurs at
energies in the range 50~keV $\le E_{cm} \le$
400~keV~\cite{Ser04}. At higher energies, this reaction has been
studied carefully in direct kinematics: at energies above 1~MeV by
Robertson {\it et al.} \cite{Rob81}, and by Mohr {\it et al.}
\cite{Moh94} in the energy range around the dominant $3^{+}$
resonance at $E_{cm}=0.711$~MeV. At BBN energies, however, direct
measurements are difficult due to extremely low cross sections
(about 29~pb at $E_{cm}=100$~keV). An attempt by Cecil {\it et
al.}~\cite{Cec96} at $E_{cm}=0.053$~MeV yielded only an upper
limit for the $S$-factor of $2.0\times10^{-7}$ MeV b which is more
than an order of magnitude higher than present estimates. A
straightforward solution to overcome this problem is offered by
the indirect method of Coulomb dissociation (CD). As will be shown
below, the dominant multipolarity involved is $E$2. When
bombarding a high-$Z$ target like $^{208}$Pb with a medium-energy
($\approx$ 150 $A$ MeV) $^6$Li beam, an intense flux of virtual
$E$2 photons is created that dissociates $^6$Li into $^2$H and
$\alpha$ with a greatly enhanced cross section. From the
energy-differential CD cross section, the radiative-capture one
can be calculated easily \cite{Bau86} provided that the
multipolarity of the respective transition is known and that
higher-order electromagnetic or nuclear contributions can be
either ignored or taken into account quantitatively.

Kiener {\it et al.} \cite{Kie91} have investigated the \zdag\
reaction by means of the CD method employing 26~$A$~MeV ${}^{6}$Li
projectiles breaking up into D+$\alpha$ in the Coulomb field of a
$^{208}$Pb nucleus. Referring to a theoretical paper by Shyam {\it
et al.}~\cite{Shy91}, Kiener {\it et al.} have claimed that their
measurement is largely free from nuclear background (the same
assumption was made in a later reevaluation of the same
dataset~\cite{Iga00}). While the astrophysical $S$-factor derived
in Ref.~\cite{Kie91} seems to agree well with theoretical
predictions at and above the resonance, a puzzling result emerged
below the resonance: the experimental data suggest a rather
constant $S$-factor in the astrophysically interesting region
below 400 keV;  most theoretical curves, however, drop with
decreasing energy \cite{{Kha98}}. As we will show in the present
paper, it is likely that this constant $S$-factor is due to
nuclear processes that cannot be ignored.

We report in this article on a new break-up measurement performed
at the SIS-18 heavy-ion synchrotron at GSI (Helmholtzzentrum f\"ur
Schwerionenforschung in Darmstadt, Germany) using a higher-energy
(150 $A$ MeV) $^6$Li beam. This higher beam energy should have
several advantages compared to the one used by Kiener {\it et
al.}: (i) the stronger forward focusing allows for a more complete
angular coverage; (ii) CD should be enhanced relative to the
nuclear contribution. In addition, we have developed a
comprehensive theoretical model of electromagnetic and nuclear
break-up processes that allows to interprete the measured data in
detail. We will show below, however, that it is unfortunately not
possible to separate experimentally electromagnetic and nuclear
contributions. Nevertheless, most of the features of the measured
data can be well explained by our model, thus giving our
calculated $^2$H($\alpha,\gamma$)$^6$Li cross sections a firm
experimental basis.

\section{Theoretical predictions}

\subsection{Radiative-capture reaction}
\label{sec:radcap}

The cross section of the $^2$H($\alpha,\gamma$)$^{6}$Li reaction
at energies $E_{cm} < 1$~MeV is dominated by radiative $E$2
capture from $d$ waves in the $\alpha+^2$H channel into the
$J^{\pi}=1^{+}$ ground state of $^{6}$Li via a prominent $3^{+}$
resonance at $E_{cm}=0.711$~MeV. In comparison, $E$1 transitions
from $p$ waves to the $^{6}$Li ground state are strongly
suppressed by the isospin selection rule for $N=Z$ nuclei due to
the almost equal charge-to-mass ratio of the deuteron and the
$\alpha$ particle. Only at very low energies ($E_{cm} \le
150$~keV), the $E$1 contribution is expected to become larger than
the $E$2 capture since the penetrabilities in $p$ and $d$ waves
exhibit a different energy dependence \cite{Rob81}.

In the past, a number of different theoretical approaches were
considered for the calculation of the low-energy
$^2$H($\alpha,\gamma$)$^{6}$Li capture cross section, see
\cite{Kha98} and re\-fe\-ren\-ces therein. They comprise, e.g.,
simple potential models and microscopic cluster models using the
re\-so\-na\-ting group method (RGM) or the generator coordinate
method (GCM). Provided that the parameters of these models are
well fitted to observable quantities like the binding and
resonance energies in the ${}^{6}$Li system and that the
asymptotic form of the bound state wave function is correctly
taken into account, good agreement between the predictions for the
low-energy cross section is found. This shows that the radiative
capture at energies below the $3^{+}$ resonance is essentially an
extranuclear process and that details of the interior wave
function are less important.

\begin{figure}[t]

\begin{center}

\includegraphics[width=1.00\linewidth]{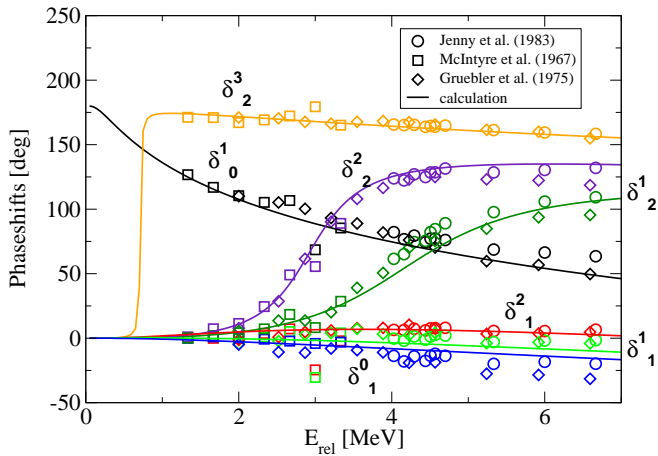}

\end{center}

\caption{(Color online) Phase-shift data measured for low-energy
$\alpha$-$^2$H scattering as a function of the relative
$\alpha$-$^2$H energy in the c.m. system, $E_{rel}$. Data points
are from Jenny {\it et al.} (circles, \protect\cite{Jen83}),
McIntyre {\it et al.} (squares, \protect\cite{McI67}), and
Gr\"uebler {\it et al.} (diamonds, \protect\cite{Gru75}). The
results of the model calculations (full lines) were obtained with
the potential parameters described in the text.}
\label{fig01_phaseshifts}

\end{figure}

In the present work we employ a potential model for the ${}^{6}$Li
system that provides the $S$-factor for the
$^2$H($\alpha,\gamma$)$^{6}$Li reaction and, at the same time, can
be used in modelling the breakup reaction when ${}^{6}$Li is
scattered on a Pb nucleus at $150~A$~MeV. This model assumes that
${}^{6}$Li is described by two interacting clusters, $\alpha$ and
$^2$H, without internal structure. Bound and scattering wave
functions in the relevant partial waves, characterized by orbital
angular momentum $l$ and total angular momentum $J$, are obtained
by solving the appropriate radial Schr\"{o}dinger equation with
$\alpha$-$^2$H potentials that contain a central part of
Woods-Saxon form
\begin{equation}
 V_{c}^{l,J}(r) = - \mathcal{V}_{c}^{l,J} \left[
 1 + \exp\left( \frac{r-R}{a}\right) \right]^{-1}
\end{equation}
and a spin-orbit part of derivative Woods-Saxon form
\begin{equation}
 V_{so}^{l,J}(r) = \mathcal{V}_{so}^{l,J} \lambda^{2}
 \frac{\vec{L}\cdot\vec{S}}{\hbar^{2} r}
 \frac{d}{dr}\left[ 1 + \exp\left( \frac{r-R}{a}\right)  \right]^{-1}
\end{equation}
with $\lambda = 2$~fm. The radius $R$ is given by
$R=1.25~A^{1/3}$~fm with $A=6$; $a=0.65$~fm denotes the
diffuseness parameter. The depths $\mathcal{V}_{c}^{l,J}$ were set
to 60.712 MeV for the ground state ($l=0$) and 56.7 MeV for all
other partial waves ($l=1,2$). For the spin-orbit part
$\mathcal{V}_{so}^{l,J}$ a depth of 2.4 MeV was used for the
relevant partial waves $l=1,2$. These values were obtained by
adjusting the parameters as to reproduce the experimental values
for the binding energy $E_{B} = 1.474$~MeV of ${}^{6}$Li and the
$3^{+}$ resonance energy with respect to the $\alpha$+$^2$H
threshold. This choice of parameters also describes the low-energy
$\alpha$-$^2$H experimental scattering phase shifts very well, see
Fig.~\ref{fig01_phaseshifts}.
The cross section of the radiative capture reaction is calculated
in the present model with the usual long-wavelength approximation
of the $E$1 and $E$2 multipole operators
$\mathcal{M}(E\lambda\mu)= Z_{\rm eff}^{(\lambda)} e r_{\alpha
D}^{\lambda} Y_{\lambda\mu}(\hat{r}_{\alpha d})$ where $e$ denotes
the electron charge and $\vec{r}_{\alpha d}$ is the radius vector
between $\alpha$ and deuteron. The effective charge numbers
\begin{equation}
 Z_{\rm eff}^{(\lambda)} =  Z_d \left(
 \frac{m_{\alpha}}{m_d+m_{\alpha}}\right)^{\lambda}  + Z_{\alpha} \left( - \frac{m_d}{m_d+m_{\alpha}}\right)^{\lambda}
\end{equation}
depend on the charge numbers $Z_{i}$ and masses $m_{i}$ of the two
clusters. The $E$1 effective charge number does not vanish since
experimental values for the masses are used. However, the $E$2
contribution dominates over most of the range of energies with a
pronounced peak at the position of the $3^{+}$ resonance. Only at
energies below 110~keV the $E$1 contribution exceeds the $E$2
contribution. We display the energy dependences of the two
relevant multipole contributions to the $S$-factor below in Fig.
\ref{fig09_sfactor} of Sect.IV.

\subsection{Breakup reaction}
\label{sec:breakup}

The theoretical description of the breakup reaction
${}^{208}$Pb(${}^{6}$Li,$\alpha ^2$H)${}^{208}$Pb is considerably
more involved than that of the radiative capture reaction, in
particular if both electromagnetic and nuclear breakup have to be
included. The differential breakup cross section in the
${}^{6}$Li-${}^{208}$Pb c.m.\ system can be written in the general
form
\begin{eqnarray}
\label{eq:buxs}
\lefteqn{\frac{d^{3}\sigma}{d\Omega_{LiPb}dE_{\alpha
 d}d\Omega_{\alpha d}}}
 \\ & = & \nonumber
 \frac{\mu_{LiPb}^{2}}{(2\pi)^{2}\hbar^{4}}
 \frac{p_{LiPb}^{f}}{p_{LiPb}^{i}}  \frac{1}{2J_{Li}+1} \sum_{M_{Li}}
 \sum_{M_{d}} \left| T_{fi} \right|^{2} \frac{\mu_{\alpha d}p_{\alpha d}}{(2\pi\hbar)^{3}}
\end{eqnarray}
with reduced masses $\mu_{ij} = m_{i}m_{j}/(m_{i}+m_{j})$ and
relative momenta $\vec{p}_{ij} = \mu_{ij} \left( \vec{p}_{i}/m_{i}
-   \vec{p}_{j}/m_{j}\right)$. $\Omega_{ij}$ denotes the solid
angle for the scattering of particles $i$ and $j$ in their c.m.\
system and $E_{\alpha d} = p_{\alpha  d}^{2}/(2\mu_{\alpha d})$ is
the c.m. energy in the fragment system after the breakup. In the
initial state, ${}^{6}$Li is in the ground state with total
angular momentum $J_{Li}=1$ and $M_{Li}=\pm 1,0$. In the final
state, the deuteron carries spin $1$ with projections $M_{d} = \pm
1, 0$. The cross section (\ref{eq:buxs}) determines the relative
probability to find the two fragments with given momenta in the
final state und thus can be used directly in a Monte-Carlo
simulation of the breakup reaction.

The main task is to calculate the T-matrix element
that contains all the relevant information on the breakup process.
In distorted wave Born approximation (DWBA) it is given by
\begin{eqnarray}
\label{eq:Tmat}
 T_{fi} & = &  \langle \chi^{(-)}(\vec{p}_{LiPb}^{\,f}) \Psi_{\alpha
  d}^{(-)}(\vec{p}_{\alpha d} M_{d})|
 \\ \nonumber & &
 \times \left( V_{LiPb} - U_{LiPb} \right) |
 \Phi_{Li}(J_{Li}M_{Li}) \chi^{(+)}(\vec{p}_{LiPb}^{\,i})\rangle
\end{eqnarray}
with the ${}^{6}$Li ground state wave function $\Phi_{Li}$ and the
wave function $\Psi_{\alpha d}^{(-)}$ for the relative motion of
the fragments in the continuum. These two functions are given by
the solutions of the Schr\"{o}dinger equation as in the
calculation of the radiative-capture cross section. The distorted
waves $\chi^{(\pm)}$ describe the scattering of the projectile on
the target. They can be found by solving the Schr\"{o}dinger
equation for the Li-Pb scattering with the optical potential
$U_{LiPb}$ that only depends on the distance between Li and Pb. In
contrast, $V_{LiPb}$ is the full many-body interaction potential.
It is approximated by
\begin{eqnarray}
 V_{LiPb} & \approx & \frac{Z_{\alpha} Z_{Pb}
   e^{2}}{\left|\vec{r}_{\alpha}-\vec{r}_{Pb}\right|}
 +  \frac{Z_{d} Z_{Pb}
 e^{2}}{\left|\vec{r}_{d}-\vec{r}_{Pb}\right|}
 \\ \nonumber & &
 + U_{\alpha Pb}^{N}(\left|\vec{r}_{\alpha}-\vec{r}_{Pb}\right|)
 + U_{d Pb}^{N}(\left|\vec{r}_{d}-\vec{r}_{Pb}\right|)
\end{eqnarray}
separating Coulomb and nuclear contributions and introducing
nuclear optical potentials $U_{\alpha Pb}^{N}$ and $U_{d Pb}^{N}$
for the $\alpha$-Pb and $^2$H-Pb interaction, respectively.
Similarly we have
\begin{equation}
\label{eq:ULiPb}
 U_{LiPb}(\vec{r}_{LiPb})  \approx
 \frac{Z_{Li} Z_{Pb} e^{2}}{\left|\vec{r}_{LiPb}\right|}
 + U_{Li Pb}^{N}(\left|\vec{r}_{LiPb}\right|)
\end{equation}
with $\vec{r}_{LiPb} = \vec{r}_{Li}-\vec{r}_{Pb}$. Since both
potentials contain Coulomb and nuclear contributions additively,
it is possible to separate the T-matrix element into a Coulomb and
a nuclear part as
\begin{equation}
 T_{fi} = T_{fi}^{C} + T_{fi}^{N} \: .
\end{equation}
In general, Coulomb and nuclear contributions to the breakup
amplitude can interfere.

In the breakup experiment, the projectile velocity $v_{LiPb}$
relative to the target is large and the fragments are observed at
small forward scattering angles with res\-pect to the beam axis.
Thus it is sufficient to replace the distorted waves appearing in
Eq.\ (\ref{eq:Tmat}) by their eikonal approximation, i.e.\
\begin{eqnarray}
 \lefteqn{\chi^{(-)\ast}(\vec{p}_{LiPb}^{f})
 \chi^{(+)}(\vec{p}_{LiPb}^{i})}
 \\ \nonumber & = &
  \exp\left[i \vec{q} \cdot \vec{b} \right]
 \exp \left[ iS_{LiPb}(\vec{b})\right]
\end{eqnarray}
with the momentum transfer
\begin{equation}
 \vec{q} = \frac{1}{\hbar} \left(
   \vec{p}_{LiPb}^{i}-\vec{p}_{LiPb}^{f} \right)
\end{equation}
and the phase function
\begin{equation}
\label{eq:phase}
 S_{LiPb}(\vec{b}) = -\frac{1}{\hbar v_{LiPb}}
 \int_{-\infty}^{\infty} dz \: U_{LiPb}(\vec{r}_{LiPb})
\end{equation}
where the coordinate vector has been decomposed as
\begin{equation}
 \vec{r}_{LiPb} = \vec{b} + z \hat{e}_{\rm beam} \qquad
 \vec{b} \perp \hat{e}_{\rm beam}
\end{equation}
and $\hat{e}_{\rm beam}$ denotes the beam direction. The Coulomb
part of the phase function can be calculated analytically. In
order to avoid a divergent result at small impact parameters, $b$,
the Coulomb potential of a point-like target charge in Eq.\
(\ref{eq:ULiPb}) is replaced by that of a homogeneous sphere. In
the eikonal approximation, the Coulomb and nuclear T-matrix
elements can be written as
\begin{equation}
\label{eq:TmatNC}
 T_{fi}^{C/N} = \langle  \Psi_{\alpha d}^{(-)}(\vec{p}_{\alpha d}
 M_{d})| F_{C/N}(\vec{r}_{\alpha d}) | \Phi_{Li}(J_{Li}M_{Li})\rangle
\end{equation}
with the Coulomb form factor
\begin{eqnarray}
\label{eq:FC}
 \lefteqn{F_{C}(\vec{r}_{\alpha d}) =}
 \\ \nonumber & &
 Z_{Pb} e \int d^{3}r_{LiPb} \:
  \exp\left[i \vec{q} \cdot \vec{b} \right]
 \exp \left[ iS_{LiPb}(\vec{b})\right]
 \\ \nonumber & & \times
 \left( \frac{Z_{\alpha}e}{\left|\vec{r}_{\alpha Pb}\right|}
+  \frac{Z_{d}e}{\left|\vec{r}_{dPb}\right|} -
\frac{Z_{Li}e}{\left|\vec{r}_{LiPb}\right|} \right)
\end{eqnarray}
and the nuclear form factor
\begin{eqnarray}
 \lefteqn{F_{N}(\vec{r}_{\alpha d})=}
 \\ \nonumber & &
 \int d^{3}r_{LiPb} \:
  \exp\left[i \vec{q} \cdot \vec{b} \right]
 \exp \left[ iS_{LiPb}(\vec{b})\right]
 \\ \nonumber & & \times
 \left( U^{N}_{\alpha Pb} + U^{N}_{dPb} - U^{N}_{LiPb}\right)
\end{eqnarray}
that can be both decomposed into multipoles $L=0,1,2,\dots$. Using
the method of steepest descent, the multipole components of the
Coulomb form factor (\ref{eq:FC}) are easily calculated.
Neglecting the nuclear contribution in the phase function
$S_{LiPb}$ one obtains in lowest order the well-known Coulomb
excitation functions in the semiclassical approximation. The
nuclear optical potential $U^{N}_{LiPb}$ in (\ref{eq:phase}) leads
to corrections that take, e.g., the absorption by the target
nucleus into account; also, relativistic corrections are easily
included, see Ref.~\cite{Typ07} for details.

Since the $E$2 virtual photons are orders of magnitude more
abundant than the $E$1 ones, the Coulomb contribution to the
breakup is essentially sensitive only to the quadrupole
contribution. (There is no monopole contribution in this case.)
Contrary to electromagnetic $E$1 excitations, nuclear $L=1$
excitations are not suppressed by isospin selection rules. Nuclear
processes for all multipolarities have, therefore, to be taken
into account when modelling the break-up of $^{6}$Li into
$\alpha+^2$H at about 150 $A$ MeV. In the present work, we
included nuclear $L=0,1,2$ excitations because higher multipoles
are expected to give only small contributions to the total breakup
amplitude.

In order to obtain numerical results for the T-matrix elements,
one has to specify the nuclear optical potentials that enter into
the calculation. Unfortunately, there are no systematic optical-
model potentials available des\-cri\-bing the elastic scattering
of $\alpha$, $^2$H and ${}^{6}$Li on a Pb target at $150~A$~MeV.
Therefore, we generated the optical potentials from systematic
optical-model potentials for nucleon-Pb elastic scattering and
folded them with the matter distribution of the projectile and the
fragments, respectively. These potentials were tuned to reproduce
published elastic-scattering data at incident energies as close to
150 $A$ MeV as possible by multiplying the real and imaginary
parts by scaling factors not too far from unity. Literature data
have been used for the elastic scattering of $^2$H + $^{208}$Pb at
55 and 70 $A$ MeV~\cite{Bet93,Oka98}, of $\alpha$ + $^{208}$Pb at
120 $A$ MeV and 175 $A$ MeV~\cite{Bon85}, and of $^6$Li +
$^{208}$Pb at 100 $A$ MeV~\cite{Schw00}.


\begin{figure}[t]
\begin{center}
\includegraphics[width=0.7\linewidth]{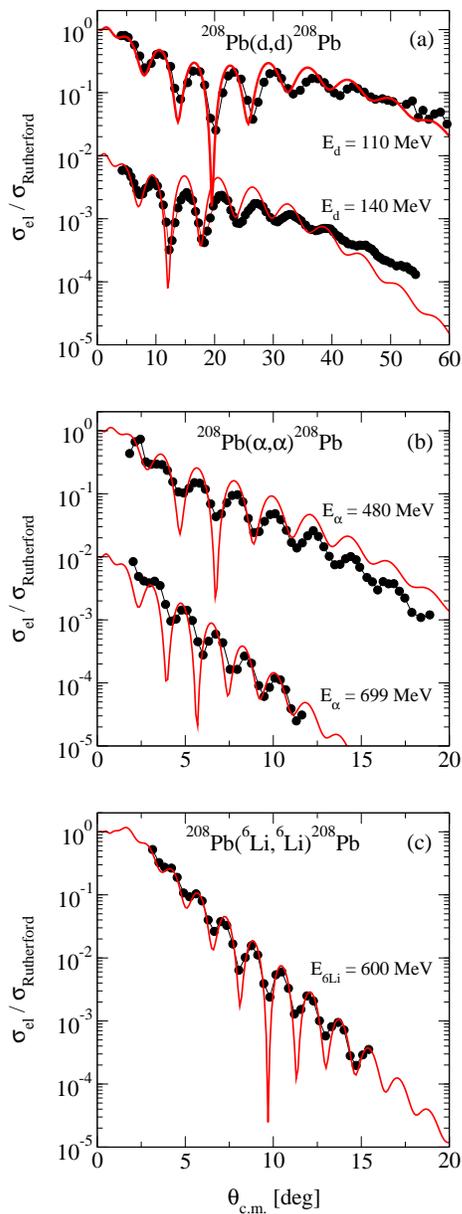}
\end{center}
\caption{(Color online) Center-of-mass angular distributions for
(a) 55 and 70 $A$ MeV $^2$H, (b) 120 and 175 $A$ MeV $\alpha$, and
(c) 100 $A$ MeV $^6$Li on $^{208}$Pb. The full lines represent
fits to the measured data using the optical-model potentials as
described in the text. Note that the angular distributions for
$E_d = 140$ MeV in panel (a) and for $E_{\alpha} = 699$ MeV in
panel (b) have been scaled by a factor of $10^{-2}$.}
\label{fig02_elastic}
\end{figure}


We found that deuteron and $\alpha$ scattering on Pb were best
described starting with the relativistic nucleon-nucleus
potentials of Ref.~\cite{CHC93}. In the case of ${}^{6}$Li-Pb
scattering the non-relativistic optical-model potential from
Ref.~\cite{BDG98} for nucleon-nucleus scattering worked best.
Fig.~\ref{fig02_elastic} shows measured and fitted
elastic-scattering data for the three cases. The optical-model
potentials, obtained by the procedure described above for the
actual energy of the breakup experiment, are well fitted by a
Woods-Saxon shape. Since mostly the outer region of the potential
is important the fits were started at a radius of 7~fm. In Table
\ref{tab:01} we give the numerical values of the depth, radius and
diffuseness parameters for the real and imaginary parts.

\begin{table}[h]
\caption{Woods-Saxon potential parameter used to describe the
scattering of $^{6}$Li, $\alpha$, and $^2$H on a Pb target.}
\label{tab:01}
\begin{tabular}{llccc}
\hline
 System         & & $^{6}$Li+$^{208}$Pb & $\alpha$+$^{208}$Pb & $^2$H+$^{208}$Pb  \\
\hline
V$_{real}$ &[MeV] & 55.0407             & 48.0315             & 23.6250       \\
R$_{real}$ &[fm]  & 7.4979              & 7.9014              & 7.9057        \\
a$_{real}$ &[fm]  & 0.8665              & 0.8542              & 0.8984        \\
V$_{imag}$ &[MeV] & 84.1720             & 45.4504             & 28.3867       \\
R$_{imag}$ &[fm]  & 7.3633              & 7.3763              & 7.3712        \\
a$_{imag}$ &[fm]  & 0.8693              & 0.9020              & 0.9391        \\
\hline
\end{tabular}\\[2pt]
\end{table}

\subsection{Predicted observables}

\begin{figure}[ht]
\begin{center}
\vspace{10mm}
\includegraphics[width=1.00\linewidth]{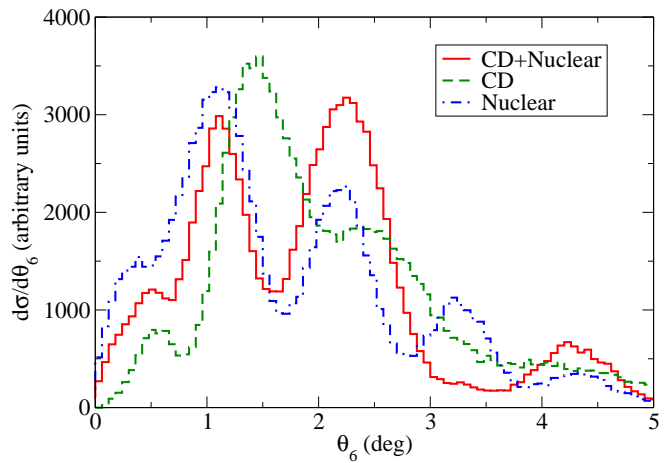}
\end{center}
\vspace{-5mm} \caption{(Color online) Expected distribution of the
differential cross section, $d\sigma/d\theta_6$, as a function of
the scattering angle, $\theta_6$, of the excited $^6$Li$^*$ before
break-up, in arbitrary units. The full (red) curve represents the
total distribution, whereas the nuclear and Coulomb contributions
are depicted by the dot-dashed (blue) and dashed (green)
histograms, respectively. Note that the different curves have been
normalized to the same total cross section. All distributions were
summed over $^2$H-$\alpha$ c.m. energies, $E_{rel}$, up to 1.5
MeV.} \label{fig03_dsdtheta}
\end{figure}

The most meaningful observable that can illustrate the predictions
from the above-sketched model of \six\ break-up is the scattering
angle, $\theta_6$, of the excited $^6$Li$^*$ before break-up,
relative to the incoming \six\ beam. Fig.~\ref{fig03_dsdtheta}
depicts the expected $\theta_6$ distribution.

The figure clearly shows that pure nuclear, pure Coulomb, and
total (CD+nuclear) distributions exhibit distinctly different peak
structures. Pure Coulomb interaction has its most prominent peak
where the other contributions show a minimum. Likewise, the total
(CD+nuclear) distribution can be distinguished from a nuclear-only
theory by the large amplitude of the most prominent peak (due to
constructive CD-nuclear in\-ter\-fe\-ren\-ce), and by the
disappearance of the third maximum (due to destructive
interference). In principle, these features should allow to
separate the contributions from the individual interactions.
However, the theoretical predictions have to be folded with the
resolution and the acceptance of the experimental apparatus using
the Monte-Carlo simulations described below in
subsection~\ref{sec:monte-carlo}.

\section{Experimental procedure}
\subsection{Apparatus}
\begin{figure*}[ht]
\begin{center}
\includegraphics[width=1.0\linewidth]{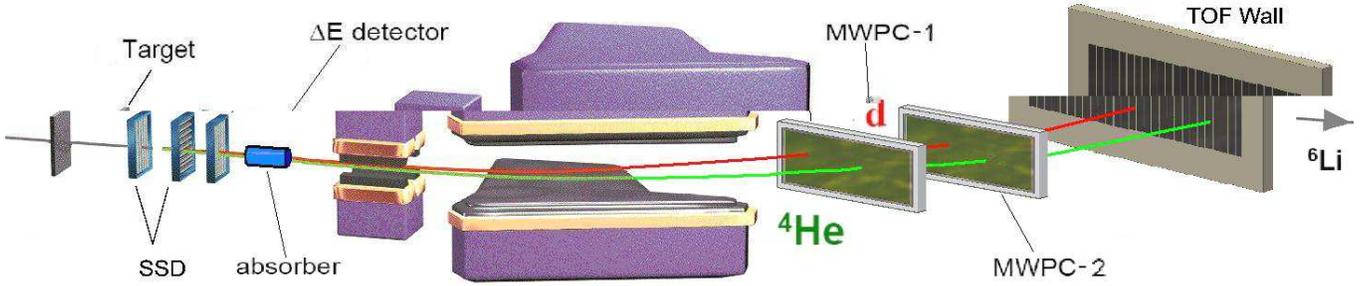}
\end{center}
\vspace{-0.5 cm} \caption{(Color online) The experimental setup
shows the fragment-tracking SSD behind the Coulomb-break-up target
followed by a 16-strip $\Delta$E detector and a beam stopper.
Deuteron and $\alpha$ positions were measured near the focal plane
of the KaoS QD-spectrometer by two successive large area
multi-wire proportional chambers (MWPC) followed by a
scintillator-paddle TOF wall used for trigger purposes.}
\label{fig04_kaos_setup}
\end{figure*}

A schematic view of the set-up used is given in
Fig.~\ref{fig04_kaos_setup}. A $^{208}$Pb target with 200
mg/cm$^2$ thickness was bombarded by a primary $^6$Li beam of 150
$A$ MeV energy. The $^6$Li beam was produced by the SIS-18
synchrotron at GSI, separated from possible contaminant ions by
using the FRS FRagment Separator~\cite{Gei92} and transported to
the standard target position of the kaon spectrometer
KaoS~\cite{Sen93}. The average $^6$Li beam intensity at the
break-up target was of the order of 5x10$^4$ per 4 sec spill. The
beam had a width of 0.17(0.12) cm and an angular divergence of
4.4(4.4) mrad in x(y) direction at the target (1$\sigma$ widths).

The angles and positions as well as the energy losses of the
outgoing particles, $^2$H and $\alpha$, were measured by two pairs
of single-sided silicon strip detectors (SSD, 300 $\mu$m thick,
100 $\mu$m pitch) \cite{Iwa99} placed at distances of 15 and 30
cm, respectively, downstream from the target. From the detector
pitch one can calculate a resolution of the $^2$H-$\alpha$ opening
angle in the laboratory, $\theta_{24}$, of about 1\%.
Non-interacting $^6$Li beam particles were identified event by
event with a 16-strip $\Delta$E detector located directly behind
the SSD and stopped in a cylindrical Ta absorber (12 mm
dia\-me\-ter, 20 mm length) placed behind the detector. Break-up
events were discriminated from non-interacting $^6$Li-beam events
by their energy-loss signals in the 16-strip $\Delta$E detector;
an energy loss corresponding to $^6$Li was used as a trigger veto
signal. Deuteron and $\alpha$ momenta were analyzed with the
large-acceptance KaoS spectrometer and were detected in two
consecutive multi-wire proportional chambers (MWPC~\cite{Iwa99})
followed by a plastic-scintillator TOF wall consisting of 30
elements (each 7 cm wide and 2 cm thick). This plastic wall was
used as a trigger detector for the data acquisition system. The
KaoS magnets' volume was filled with He gas at atmospheric
pressure to reduce multiple scattering.

The coincident $^2$H and $\alpha$ fragments resulting from
break-up in the $^{208}$Pb target were identified by
reconstructing their vertex at the target. This removed all
break-up events in layers of matter other than the target. The
$^2$H and $\alpha$ momenta were determined from tracking them with
GEANT through the MWPC and TOF wall behind KaoS. The incident
angles in front of the magnets were known unambiguously from the
SSD hits. While in the SSD each hit could be attributed to either
$^2$H or $\alpha$ by its energy deposition, the corresponding hits
in the MWPC were attributed to the respective particle type by
finding the optimum trajectory through the MWPC and the TOF wall.
This was done in an i\-te\-ra\-ti\-ve procedure that started with
a test assignment of each hit to either $\alpha$ or $^2$H and a
test momentum for each of them. Both the momentum values and the
assignments were then iteratively changed until the minimum
squared deviation from the observed hits in all detectors
downstream from the KaoS magnet were reached. This momentum
reconstruction could be shown to be accurate within about
$10^{-3}$. From the opening angles between the fragments and from
their momenta, the relative energies, $E_{rel}$, between the $^2$H
and $\alpha$ particles in the c.m. system could be reconstructed.

\subsection{Monte-Carlo simulations}
\label{sec:monte-carlo}

It is obvious that the experimental apparatus imposes strong
restrictions on the detection of the break-up particles, $\alpha$
and $^2$H. This applies in particular to the angular acceptance,
the energy and position resolution, and the detection efficiency.
As a consequence, a meaningful comparison between theoretical
predictions and experimental data can only be made using
theoretical data filtered by the experimental set-up. To this end,
we have modelled the entire set-up, starting in front of the
$^{208}$Pb break-up target, in GEANT3~\cite{geant3}. As an event
generator, the theoretical model described in the previous section
was used. Input data were generated as statistically-distributed
ensembles of 100,000 break-up``events'' each that were distributed
according to the calculated differential cross sections. The
emittance of the $^6$Li beam (as measured without break-up target
and without absorber) was imposed. Each break-up particle,
$\alpha$ and $^2$H, was followed through the remainder of the Pb
target after the reaction vertex, the SSD detectors, the beamline
exit window, the He-filled interior of the magnets with the
magnetic field and the air behind KaoS before hitting the MWPC
volumes and the TOF wall.

\begin{figure}[h]
\begin{center}
\vspace{-8mm}
\includegraphics[width=0.85\linewidth]{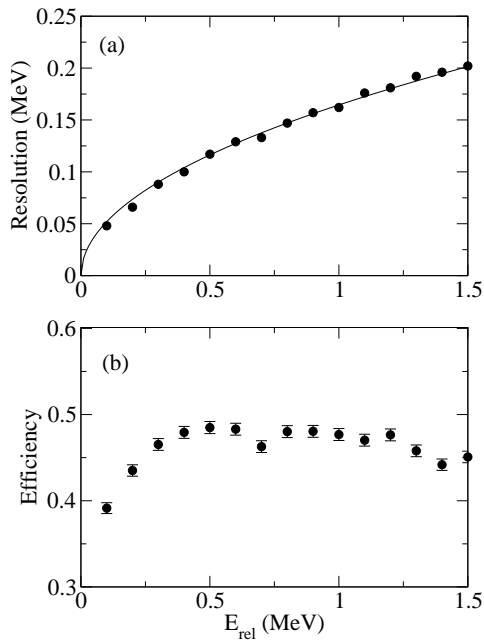}
\end{center}
\vspace{-0.5 cm} \caption{(a) Relative-energy resolution
(1$\sigma$ widths) as determined by simulating $^6$Li breakup with
GEANT. The data points can be approximated by the fitted function
$\sigma_E = 0.1645 \times \sqrt{E_{rel}}$. (b) Combined
geometrical and analysis efficiency of determining $E_{rel}$ from
the $^2$H and $\alpha$ momentum vectors. The intrinsic efficiency
of the MWPC detectors has been assumed to be unity in this graph.}
\label{fig05_res_effic}
\end{figure}

The Monte-Carlo simulations were used to obtain estimates of the
resolution and the efficiency of our setup. As an example, we plot
in Fig.\ref{fig05_res_effic}(a) the  1$\sigma$-resolution of
$E_{rel}$. The data points were obtained by sending 10,000 events
each with different values of $E_{rel}$ (within a narrow bin of
0.1 MeV width for each case) into our setup and analyzing the
outgoing particles with the same routines as in the experiment.
From the same data sets, the number of counts gave an approximate
estimate of the detection efficiency, shown in
Fig.\ref{fig05_res_effic}(b). In the experiment, however, the
detection efficiency is additionally li\-mi\-ted by the small and
strongly fluctuating energy deposition of deuterons in the MWPC.
This latter quantity cannot be simulated easily, such that we had
to normalize the number of observed and simulated counts.
Therefore, our experiment does not allow to determine absolute
cross sections, despite the fact that all incident $^6$Li ions
were counted.

\section{Experimental results}

\begin{figure}[ht]
\begin{center}
\vspace{7mm}
\includegraphics[width=1.0\linewidth]{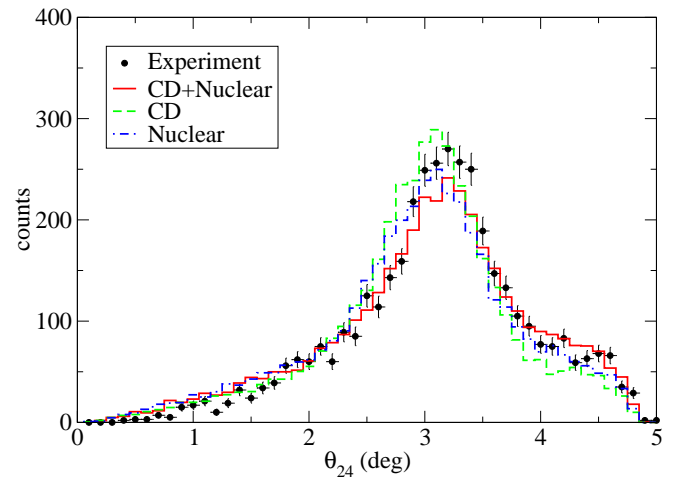}
\end{center}
\vspace{-0.5 cm} \caption{(Color online) Opening angles,
$\theta_{24}$, between the outgoing fragments $^2$H and $\alpha$.
Full circles correspond to measured data. The dash-dotted
histogram (blue) denotes simulations with pure nuclear
interaction, whereas the pure CD contribution is shown by the
dashed histogram (green). Combined (CD+nuclear) contributions are
shown by the full red line. Note that the numbers of simulated
counts in each spectrum were normalized to the experimental ones.}
\label{fig06_theta24}
\end{figure}

A relatively unbiased observable, based only on high-resolution
SSD measurements, is the opening angle, $\theta_{24}$, between the
outgoing fragments $^2$H and $\alpha$. Fig.~\ref{fig06_theta24}
shows this distribution, summed over $E_{rel}$ values up to 1.5
MeV; this condition was also set for all other spectra shown
below. The experimental data points are compared with the
corresponding Monte-Carlo simulations for pure Coulomb (CD) and
pure nuclear interactions as well as combined (CD+nuclear)
interactions. Each simulated histogram was normalized to contain
the same number of counts as the experimental spectrum, thus
providing the single scaling factor used to normalize all
simulated distributions.

As seen in Fig.~\ref{fig06_theta24}, the data are reasonably well
reproduced by the simulations over their entire range. The 3$^+$
resonance peak is clearly visible around 3 degrees; its angular
width is well reproduced indicating that the simulation takes both
the scattering and the finite angular resolution well into
account. It is obvious, however, that one cannot distinguish
between the different interactions on the basis of this angular
distribution. We have therefore to search for an observable that
is more sensitive to the type of interaction. In
Fig.~\ref{fig03_dsdtheta} above, we have shown that the observable
$\theta_6$ should be very sensitive to the type of interaction.
\begin{figure}[t]
\begin{center}
\includegraphics[width=0.9\linewidth]{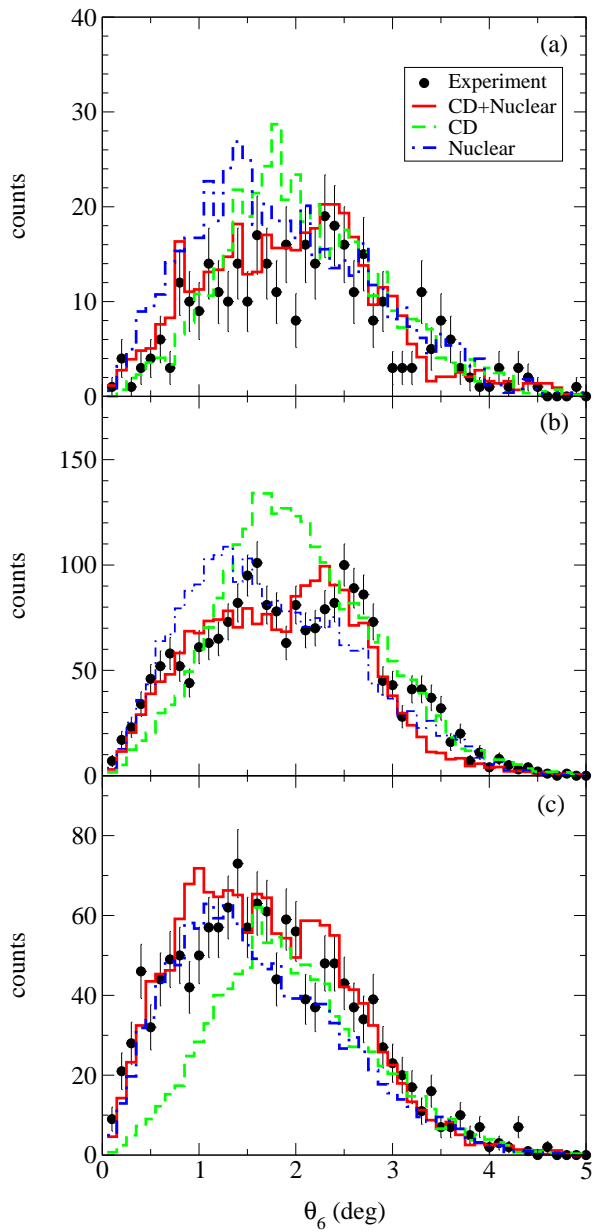}
\end{center}
\vspace{-5mm} \caption{(Color online) Angular distribution of the
excited $^6$Li$^*$ nuclei after the reaction ($\theta_6$). The
panels represent three different bins of $E_{rel}$: (a) $ 0.0 \leq
E_{rel} < 0.5$ MeV; (b) $ 0.5 \leq E_{rel} < 0.9$ MeV; (c) $ 0.9
\leq E_{rel} < 1.5$ MeV. The measured data points are shown in
comparison with simulations with pure nuclear and pure CD as well
as with (CD+nuclear) interactions. Line types and color codes are
identical to the ones in Fig.~\protect\ref{fig06_theta24}.}
\label{fig07_theta6}
\end{figure}

The experimental data for this observable are presented in
Fig.~\ref{fig07_theta6}. Panel (a) of Fig.~\ref{fig07_theta6}
shows data for $E_{rel}$ values below the resonance; panel (b)
covers the resonance region, whereas panel (c) has been plotted
for $0.9 \leq E_{rel} < 1.5$ MeV. The finite KaoS acceptance cuts
the distributions at about 4 degrees. The figure shows clearly
that the observable $\theta_6$ is sensitive to the type of
interaction. In all panels, the combined (CD+nuclear) interaction,
including interference, reproduces most of the structures observed
in the data points (red histograms). This is particularly true for
the sub-resonance region, panel (a). The green histograms
(CD-only) show single peaks at larger angles. The pure nuclear
interaction (blue histograms) rises rapidly at small values of
$\theta_6$, in agreement with the measured data, but lacks the
structures visible in the data points. The narrow peaks visible in
the on-resonance data, panel (b), at values of $\theta_6$ of
$\approx 1.5, 2.6$ and $3.3$ degrees, are not perfectly reproduced
by the (CD+nuclear) model and point to small deficiencies of the
theoretical model. Nevertheless, Fig.~\ref{fig07_theta6}
demonstrates that Coulomb-nuclear interference is at work and that
the signs of the interference terms are correct. We also conclude
that even at our incident energy of 150 $A$ MeV the nuclear
break-up is dominant.

\begin{figure}[ht]
\begin{center}
\vspace{5mm}
\includegraphics[width=1.0\linewidth]{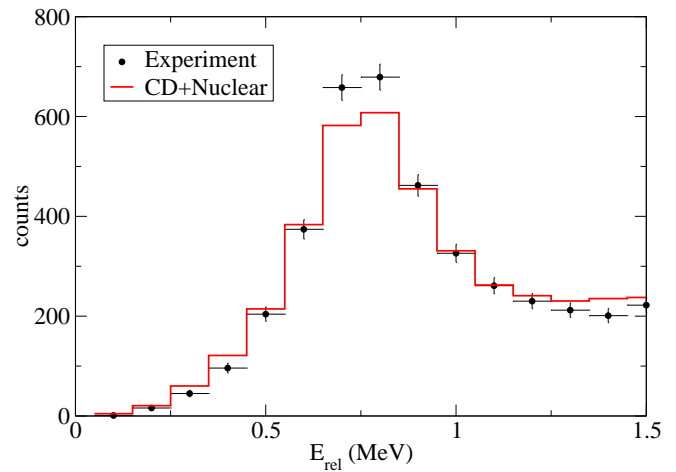}
\end{center}
\vspace{-5mm} \caption{(Color online) Differential cross sections
as a function of the energy, $E_{rel}$, in the $\alpha$-$^2$H c.m.
system. Black points indicate the experimental data; the histogram
corresponds to the GEANT simulation using the (CD+nuclear)
interaction as described in the text and a binning of 100 keV
(note that the vertical error bars result from a quadratic sum of
statistical and sys\-te\-ma\-ti\-cal uncertainties).}
\label{fig08_erel}
\end{figure}

The angle-integrated energy-differential cross sections as a
function of $E_{rel}$ are shown in Fig.~\ref{fig08_erel}.
The full histogram was obtained from the (CD+nuclear) calculation
convoluted by our GEANT simulation and normalised to the
experimental yield. The points and the histogram  represent the
measured and predicted differential cross sections, respectively,
as a function of $E_{rel}$. Our $E_{rel}$ distribution is in very
good agreement with the simulation in particular in the energy
region below 400 keV. As we will show below (see
Fig.\ref{fig10_ratio}), the differential cross sections in this
energy regime result mostly from nuclear interactions.

The astrophysically important quantity is the astrophysical
$S$-factor, $S_{24}$, for the \zdag reaction. Since nuclear
processes dominate, in particular for low $E_{rel}$, the
determination of this quantity via an evaluation of the CD
component in our break-up data is not feasible since. However, we
have demonstrated above that our theoretical model describes well
the measured cross sections, hence the astrophysical $S$-factors
from the present work are those from our theoretical model.

\begin{figure}[t]
\begin{center}
\vspace{8mm}
\includegraphics[width=0.95\linewidth]{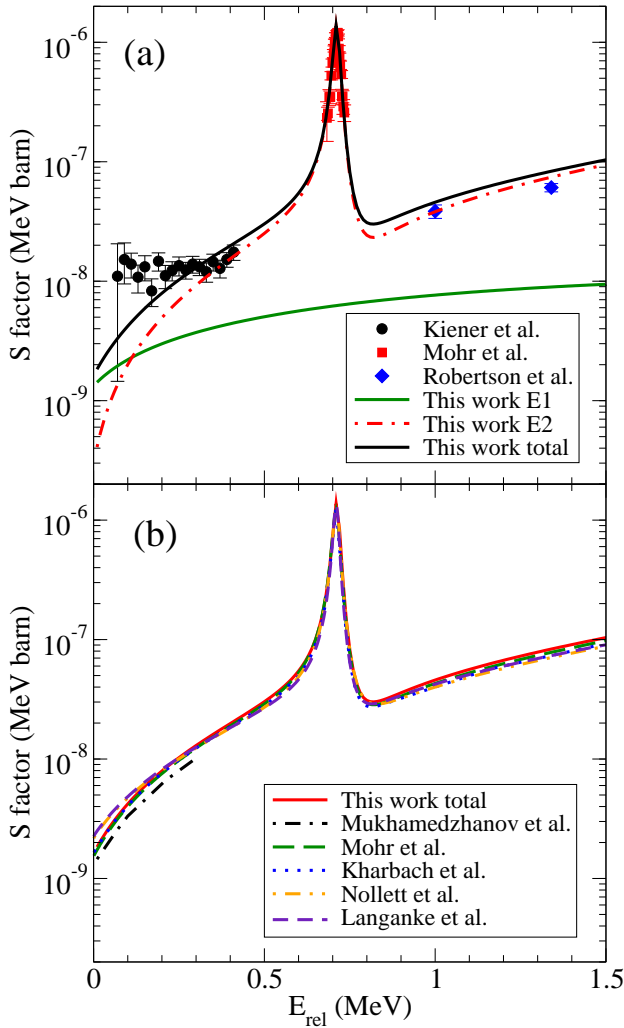}
\end{center}
\caption{(Color online) (a) Theoretical $E$1, $E$2, and total
$S_{24}$-factors that describe well the present experimental data,
together with data points from the previous CD experiment by
Kiener {\it et al.} \protect\cite{Kie91} and from direct
measurements (Robertson {\it et al.} \protect\cite{Rob81} and Mohr
{\it et al.} \protect\cite{Moh94}). See
Sect.\protect\ref{sec:comparison} for an interpretation of the
data of Ref.~\cite{Kie91}. (b) Comparison of various theoretical
predictions for the summed $E$1- and $E$2 contributions to
$S_{24}(E)$~\protect\cite{Muk95,Moh94,Kha98,Nol01,Lan86}.}
\label{fig09_sfactor}
\end{figure}

The resulting $E$1-, $E$2-, and total $S_{24}$-factors are
vi\-sua\-li\-zed in the upper part of Fig.~\ref{fig09_sfactor}
together with the previous CD data of Kiener {\it et
al.}~\cite{Kie91} and the direct data of Mohr {\it et al.}
\cite{Moh94} and Robertson {\it et al.} \cite{Rob81}.
The present results for the $E$2-component are in good agreement
with the direct measurements of Refs.~\cite{Moh94,Rob81} in the
resonance region and above which gives confidence in our model.

Another check of the validity of our treatment of the Coulomb part
of the break-up reaction, described in Sect.\ref{sec:radcap}, can
be done by comparing our calculated resonance parameters for the
$3^+$ resonance with the experimental ones. In order to determine
those parameters, we have calculated the theoretical capture cross
section around the resonance in 1 keV steps and fitted a
Breit-Wigner parametrization to the resonance. We obtain
$\Gamma$-widths of $\Gamma_{\alpha}= 22.1$ keV and
$\Gamma_{\gamma}=0.437$ meV, in good agreement with the literature
values of $\Gamma_{\alpha} = 24 \pm 2$ keV and $\Gamma_{\gamma} =
0.440 \pm 0.030$ meV as cited by Mohr {\it et al.}~\cite{Moh94}.
Note that we have used a spectroscopic factor of unity. We will
comment on the data points from the previous CD
experiment~\cite{Kie91} in the following section.

The direct \zdag reaction at very low energies is sensitive also
to the $E$1 amplitude. In our experiment, this component cannot be
constrained experimentally due to the weak flux of virtual $E$1
photons. We have therefore to rely exclusively on the theoretical
model. At higher energies, however, Robertson {\it et
al.}~\cite{Rob81} could separate $E$1- and $E$2-components on the
basis of measured angular distributions. Fig.1 in \cite{Rob81}
shows that their theory seems to overestimate the $E$1 component.
Our $E$1 curve is very close to Robertson's so that we also seem
to overestimate this component.

Several theoretical models for $^6$Li have been proposed to
determine the shape and the magnitude of the $S_{24}$ energy
dependence, such as potential models \cite{Lan86,Moh94,Typ07},
cluster-model calculations~\cite{Kha98}, or ab-initio
calculations~\cite{Nol01}. Those predicted curves for S$_{24}$
which include both $E$1- and $E$2-contributions are displayed
together with the theoretical curve from this work in the lower
panel of Fig.~\ref{fig09_sfactor}. As one can see in this figure,
all the calculations shown --independent of their very
dif\-fe\-rent model assumptions-- yield very similar curves. We
have not included the theory of Blokhintsev {\it et
al.}~\cite{Blo06}) because it was specifically tuned to approach
the experimental data of Ref.~\cite{Kie91}.

\section{Comparison with other CD experiments}
\label{sec:comparison}

\begin{figure}[ht]
\begin{center}
\vspace{5mm}
\includegraphics[width=1.0\linewidth]{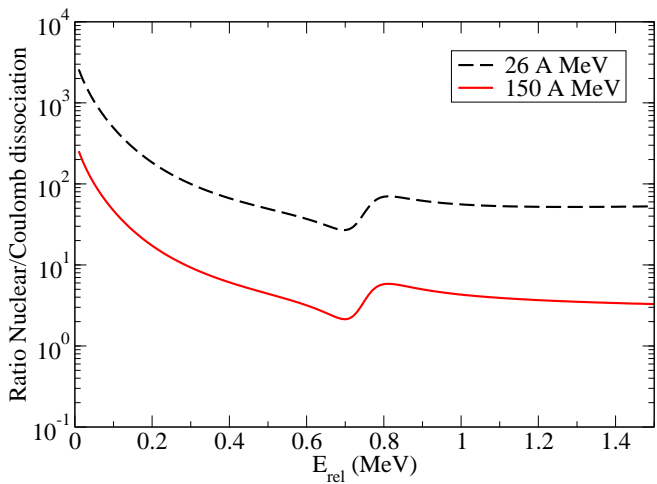}
\vspace{-5mm}
\end{center}
\caption{(Color online) Ratio of nuclear and Coulomb differential
dissociation cross sections for $^6$Li at 150 $A$ MeV (full line)
and at 26 $A$ MeV (dashed line). Both curves were calculated with
the same model described in detail above in section II.B.}
\label{fig10_ratio}
\end{figure}

As visible in the upper panel of Fig.~\ref{fig09_sfactor}, the
low-energy data points derived for the $S_{24}$-factor from the
work of Kiener {\it et al.}~\cite{Kie91} disagree with the
theoretical curve that we have deduced from the present work. We
believe that this is most likely due to a strong nuclear
contribution at the lower incident energy of 26 $A$ MeV, which was
not considered in their analysis in view of the theoretical
prediction~\cite{Shy91}. We have performed a calculation with the
theoretical model of this work at an incident energy of 26 $A$ MeV
and have calculated CD and nuclear cross sections for the
laboratory-angular range between 1.5 and 6 degrees, which should
correspond approximately to the acceptance of the setup of
Ref.~\cite{Kie91}. Fig.~\ref{fig10_ratio} displays the ratio of
nuclear to Coulomb break-up cross sections as a function of
$E_{rel}$ for the two bombarding energies. Our theory predicts
that at 150 $A$ MeV the nuclear cross sections are about a factor
of 3 larger than the CD ones at and above the resonance, whereas
the nuclear component dominates strongly at the lowest energies.
At 26 $A$~MeV, the ratio of nuclear to CD cross sections is
predicted to be about a factor of ten larger than at 150 $A$~MeV
over the entire range of $E_{rel}$. This suggests that the data
points shown in Ref.~\cite{Kie91} result almost exclusively from
nuclear interactions, contrary to the assumptions underlying their
analysis. It is therefore not very meaningful to tune theoretical
models in order to improve their agreement with the 26 $A$ MeV
data as was done in Ref.~\cite{Blo06}.

\section{Production of $^6$Li in the Big Bang}

The \zdag\ reaction is the main path for \six\ BBN production
while destruction proceeds via the $^6$Li(p,$\alpha)^3$He one.
Both rates are available in the NACRE~\citep{NACRE} compilation.
While the latter reaction rate is reasonably well known at BBN
energies, prior to this experiment the former suffered from a
large uncertainty. This was mainly due to the fact that the
published error margins were aimed at including the Kiener {\it et
al.}~\cite{Kie91} measurement~\cite{carmen}. As a result from the
present study, we can now propose a more reliable central value
based on a successful theoretical model, and a safe upper limit
that is even somewhat smaller than the previous NACRE upper limit.

\begin{figure}[tb]
\begin{center}
\includegraphics[width=\linewidth]{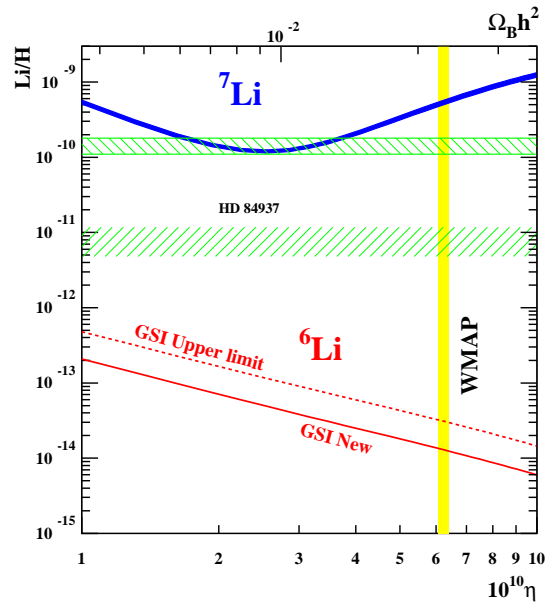}
\vspace{-8mm}
\end{center}
\caption{(Color online) Predicted BBN production ratios for
$^{6,7}$Li over hydrogen as a function of $\eta$, the
baryon-to-photon ratio in the early Universe. The solid red line
represents the result for \six\ from the $S_{24}$-values obtained
in the present work, based on theoretical values for the $E$1 and
$E$2 components. The dashed red line represents a very
conservative but safe upper limit where all observed events are
assumed to result from Coulomb break-up. The blue band denotes the
range of predicted \sep\ yields~\cite{Coc10}. Observational data
are indicated by horizontal green-hatched areas: the upper one has
been derived from the recent review of lithium observations by
Spite and Spite~\cite{Spi10}; the lower one corresponds to the
largest $^6$Li yield reported for the star HD 84937~\cite{Ste10}.
The yellow vertical band shows the WMAP $\eta$-value~\cite{WMAP}.}
\label{fig11_bigbang}
\end{figure}

In Fig.~\ref{fig11_bigbang}, the BBN abundances of \sep\ and \six\
are displayed as a function of the baryonic density. (It is usual
to introduce another parameter, $\eta$, the ratio of  the number
of baryons over the number of photons which remains constant
during the expansion, and which is directly related to \ob\ by
\obh=3.65$\times10^7\eta$). The blue \sep\ band is the result of a
Monte Carlo calculation taking into account nuclear uncertainties
as described in Ref.~\cite{Coc10}. The upper hatched horizontal
area in the figure shows the primordial lithium (\six+\sep)
abundance derived from the ``Spite plateau'', i.e. from the
practically metallicity-independent Li observations in metal-poor
stars~\cite{Spi10}.

The solid red line for \six\ has been calculated within the same
physical model, using the theoretical reaction rate from this work
for \zdag. Both $E$1 and $E$2 contributions have been included. At
WMAP baryonic density, a value for the $^6$Li/H production ratio
of $\approx 1.3 \times 10^{-14}$ results. The dashed red line
represents a very conservative upper limit for \zdag\ that would
hold if the low-energy $S_{24}$ data points from this work would
result from CD only. Fig.~\ref{fig07_theta6} demonstrates clearly
that this is not the case. But even this extremely conservative
limit is about two orders of magnitude smaller than the only
positive observation of \six\ surviving after the reanalysis of Li
lines by Steffen {\it et al.}~\cite{Ste10}, indicated in
Fig.~\ref{fig11_bigbang} by the lower hatched horizontal band.
This finding corroborates earlier statements
(e.g.~\cite{Coc10,Spi10}) that observations --if confirmed-- of
$^6$Li primordial yields around a few percent of the Spite plateau
would require astrophysical sources other than BBN.

In order to facilitate astrophysical calculations of stellar \six\
synthesis with our new theoretical $E$1 and $E$2 $S$-factors, we
list in Table~\ref{tab:02} the reaction rates for the temperature
range $10^6 {\rm K} \leq T \leq 10^{10} {\rm K}$.

\begin{table}[t]
\begin{center}
\vspace{3mm}
\begin{tabular}{||*{2}{c|}|*{2}{c|}|}
\hline
T$_9$ & N$_a\langle\sigma v \rangle$ & T$_9$ & N$_a\langle\sigma v\rangle$\\
          &  cm$^3$mol$^{-1}$s$^{-1}$ & & cm$^3$mol$^{-1}$s$^{-1}$\\
\hline
\texttt{ } 0.001 \texttt{ }  & 0.9153 $\times$ 10$^{-29}$ &\texttt{ } 0.140 \texttt{ } & 0.6967 $\times$ 10$^{-04}$ \\
0.002  & 0.2610 $\times$ 10$^{-22}$ & 0.150 & 0.9495 $\times$ 10$^{-04}$\\
0.003  & 0.3458 $\times$ 10$^{-19}$ & 0.160 & 0.1261 $\times$ 10$^{-03}$ \\
0.004  & 0.3190 $\times$ 10$^{-17}$ & 0.180 & 0.2090 $\times$ 10$^{-03}$ \\
0.005  & 0.7929 $\times$ 10$^{-16}$ & 0.200 & 0.3237 $\times$ 10$^{-03}$ \\
0.006  & 0.9163 $\times$ 10$^{-15}$ & 0.250 & 0.7846 $\times$ 10$^{-03}$ \\
0.007 & 0.7672 $\times$ 10$^{-14}$ & 0.300  & 0.1557 $\times$ 10$^{-02}$ \\
0.008 & 0.4990 $\times$ 10$^{-13}$ & 0.350  & 0.2715 $\times$ 10$^{-02}$ \\
0.009 & 0.2100 $\times$ 10$^{-12}$ & 0.400  & 0.4325 $\times$ 10$^{-02}$ \\
0.010 & 0.6547 $\times$ 10$^{-12}$ & 0.450  & 0.6453 $\times$ 10$^{-02}$ \\
0.011 & 0.1655 $\times$ 10$^{-11}$ & 0.500  & 0.9169 $\times$ 10$^{-02}$ \\
0.012 & 0.3612 $\times$ 10$^{-11}$ & 0.600  & 0.1674 $\times$ 10$^{-01}$ \\
0.013 & 0.7142 $\times$ 10$^{-11}$ & 0.700  & 0.2813 $\times$ 10$^{-01}$ \\
0.014 & 0.1325 $\times$ 10$^{-10}$ & 0.800  & 0.4502 $\times$ 10$^{-01}$ \\
0.015 & 0.2363 $\times$ 10$^{-10}$ & 0.900  & 0.6944 $\times$ 10$^{-01}$ \\
0.016 & 0.4103 $\times$ 10$^{-10}$ & 1.000  & 0.1033 $\times$ 10$^{+00}$ \\
0.018 & 0.1157 $\times$ 10$^{-09}$ & 1.250  & 0.2359 $\times$ 10$^{+00}$ \\
0.020 & 0.2965 $\times$ 10$^{-09}$ & 1.500  & 0.4350 $\times$ 10$^{+00}$ \\
0.025 & 0.2014 $\times$ 10$^{-08}$ & 1.750  & 0.6839 $\times$ 10$^{+00}$ \\
0.030 & 0.8452 $\times$ 10$^{-08}$ & 2.000  & 0.9623 $\times$ 10$^{+00}$ \\
0.040 & 0.6594 $\times$ 10$^{-07}$ & 2.500  & 0.1549 $\times$ 10$^{+01}$ \\
0.050 & 0.2827 $\times$ 10$^{-06}$ & 3.000  & 0.2132 $\times$ 10$^{+01}$ \\
0.060 & 0.8598 $\times$ 10$^{-06}$ & 3.500  & 0.2705 $\times$ 10$^{+01}$ \\
0.070 & 0.2094 $\times$ 10$^{-05}$ & 4.000  & 0.3280 $\times$ 10$^{+01}$ \\
0.080 & 0.4372 $\times$ 10$^{-05}$ & 5.000  & 0.4476 $\times$ 10$^{+01}$ \\
0.090 & 0.8156 $\times$ 10$^{-05}$ & 6.000  & 0.5754 $\times$ 10$^{+01}$ \\
0.100 & 0.1397 $\times$ 10$^{-04}$ & 7.000  & 0.7088 $\times$ 10$^{+01}$ \\
0.110 & 0.2240 $\times$ 10$^{-04}$ & 8.000  & 0.8438 $\times$ 10$^{+01}$ \\
0.120 & 0.3406 $\times$ 10$^{-04}$ & 9.000  & 0.9773 $\times$ 10$^{+01}$ \\
0.130 & 0.4959 $\times$ 10$^{-04}$ & 10.00  & 0.1107 $\times$ 10$^{+02}$ \\
\hline
\end{tabular}
\end{center}
\caption{\small Recommended $^2$H($\alpha,\gamma$)$^6$Li reaction
rates using theoretical $E1-$ and $E2-S$-factors from the present
work for the temperature range $10^6 {\rm K} \leq T \leq 10^{10}
{\rm K}$ ($10^{-3} \leq T_9 \leq 10$).}\label{tab:02}
\end{table}

\section{Conclusions}

A kinematically complete measurement of the high-energy break-up
of $^6$Li at 150 $A$ MeV has shown that Coulomb and nuclear
contributions and their in\-ter\-fe\-ren\-ces have to be taken
into account when interpreting the measured angular distributions.
Though it was not possible to extract the Coulomb part
experimentally, we were able to infer the $E$2 component of the
astrophysical $S_{24}$-factor for the
$^2$H($\alpha$,$\gamma$)$^6$Li reaction from a theoretical
reaction model that describes well in particular the low-energy
break-up data. The model predicts a drop of $S_{24}$ with
decreasing relative $^2$H-$\alpha$ energy, $E_{rel}$, as predicted
also by most other nuclear models for $^6$Li, contrary to
conclusions from an earlier CD experiment performed at the lower
energy of 26 $A$ MeV. We have presented evidence that this earlier
experiment probably has measured mostly nuclear break-up of \six.
Our findings allow to make new predictions for the $^6$Li/H
production ratio in Big-Bang nucleosynthesis (BBN) which is orders
of magnitudes smaller than the one derived from claimed
observations of \six\ in old metal-poor stars. Sources other than
BBN have therefore to be invoked for \six\ production if those
observations are confirmed.

\acknowledgments

We thank N.~Kurz for his tireless help with the data acquisition.
Thanks go to P.~Descouvemont and K.~Nollett for kindly providing
numerical results from their theo\-re\-ti\-cal calculations. This
research was supported by the DFG Cluster of Excellence ``Origin
and Structure of the Universe''.

\end{document}